# A STUDY OF A REAL-TIME OPERATING SYSTEM ON THE INTEL-BASED VME CONTROLLERS

T. Ohata and T. Masuda, SPring-8, Hyogo 679-5198, Japan

*Abstract*

We tested the real-time characteristics of varieties of VME-based CPU boards and operating systems to select the next generation controller for the SPring-8 controls. Previously we had used the HP9000/743rt, but Hewlett-Packard discontinued the model and no longer support it. We chose a VME-based Intel-architecture (IA32) CPU board and Solaris 7 as the next platform after measurements of real-time performance. The interrupt response time, the data transfer speed and the context switching time were measured as a guide of real-time performances. The IA32 platform operated with the Solaris shows good enough performance for our control system. In this paper, we report on the real-time characteristics of the operating systems, HP-RT, Solaris and standard Linux as the comparative study.

## 1 INTRODUCTION

The control system in SPring-8, a third generation light source facility, is designed on network-distributed system [1]. We used HP9000/743rt (PA-RISC PA-7100LC 64MHz) CPU boards with HP-RT version 2.21 real-time operating system as VME device controllers. The HP-RT is based on LynxOS [2]. Because Hewlett-Packard announced that the supply and support for it would be discontinued, we had to replace the CPU boards. We explored the next coming controllers and operating systems from non-proprietary systems.

First, we selected the CPU architecture. From a non-proprietary philosophy we chose the IA32 system. Today, it is possible to obtain performance and stability for the real-time control system on the IA32 platform; it has grown up to be a reliable system for large control systems.

Second, we had to select the bus architecture. We decided to keep the VME-based architecture in order to match the large amount of currently used VME-based I/O systems.

Third, we investigated the operating system for the controller. In SPring-8 controls, the compatibility with UNIX function calls is highly important for the porting of the current control system. After studying the real-time features, we chose Solaris 7 as the first priority and Linux as a substitute OS. Solaris 7 has some scheduling class such as time-sharing (TS) class to keep compatibility with standard UNIX and real-time (RT) class for real-time applications [3].

## 2 TEST ENVIRONMENTS

We tested the variety of IA32 platforms as shown in Table 1. All of the IA32 CPU boards build on the so-called PC-AT architecture. A *de facto* standard PCI-VMEbus bridge, Tundra Universe-II, is embedded on these CPU boards except PCISYS-56A, which has another bridge made by Advanet Inc. A bridge converts the different endians for adapting PCIbus and VMEbus. There is some danger that this might endanger performance of the system.

The Digital I/O (DIO) board was used for the measurement of the interrupt response time when an interrupt came from VMEbus. Also the SRAM board was used to measure the data transfer speed with the data size, D8, D16 and D32.

Table 1: Specifications of the platform on the performance measurements

| Maker | Hewlett-Packard | GMS [1] | Xycom [2] | Advanet [3] | Gespac [4] |
|---|---|---|---|---|---|
| Product | 743rt | V155 | XVME-658 | Advme8001 | PCISYS-56A |
| CPU | PA7100LC | Pentium MMX | K6-2 | PentiumIII | Pentium MMX |
| Clock(MHz) | 64 | 166 | 333 | 600 | 200 |
| Memory(MB) | 16 | 128 | 64 | 128 | 64 |
| Bus | VME | VME | VME | VME | CompactPCI |

---

[1] General Micro Systems, Inc.: www.general-micro-systems.com
[2] Xycom Automation, Inc.: www.xycom.com
[3] Gespac: www.gespac.com
[4] Advanet Inc.: www.advanet.co.jp

# 3 MEASUREMENTS

The performances of the system such as real-time characteristics are essential. We measured a part of the Rhealstone benchmark [4] of the systems by comparing HP743rt with IA32 platforms from the point of, 1) interrupt response time, 2) data transfer speed to/from VMEbus and 3) context switching time.

The interrupt response time and the data transfer speed were measured in the kernel space by using a device driver which we developed. The HP743rt, V155 and the PCISYS-56A were tested for the comparison of CPU and bus architecture dependency. Solaris 7 with TS-class was used to measure these performances on IA32 systems.

We measured the context switching time in the user space. Solaris 7 with RT-class and standard Linux 2.2.9 were tested on the IA32 system. We can fix the priority of the process for the real-time scheduling on the RT-class, which satisfies the requirements of real-time applications. HP-RT was also tested for the comparison of the real-time performance.

We used three kinds of clock references to measure the processing time. One is the *gettimeofday()* standard system call. It has a microsecond resolution but it depends on the OS implementation. The second is the interval timer (82C54) of the south bridge on the IA32 CPU board. It counts 1.193180MHz clock and has enough time resolution for comparing between IA32 systems. And the third is the 100MHz timer board on the VMEbus, which is used for comparison between the HP743rt and IA32 systems.

## 3.1 Interrupt Response Time

The interrupt response time means the delay time between an interrupt event on the hardware and the start up of the interrupt service routine of the device driver. This value includes all sources of latencies: hardware, processor dispatch, low-level interrupt handling and kernel thread dispatch. The interrupt response time, deadtime of the system influences the processing performance. To make the quick response from interrupt SIGNAL_ONLY interrupt type was used [5][6] in HP-RT.

Table 2 shows the interrupt response time with CPU and bus architecture dependency. IA32 systems are about 5-10% faster than HP743rt despite using the PCI-VMEbus bridge. This appears to show that the interrupt response time mostly depends on the CPU clock speed.

## 3.2 Data Transfer Speed

The data transfer speed means the effective throughput between the main memory of the CPU and the memory on the VME boards. We only measured programmed I/O mode (without DMA mode) in the test. The performance of the bus-bridge contributes to the data transfer speed. We measured the address space dependency of memories on the VMEbus. We disabled a posted-write mode of the Tundra Universe-II [7] to measure net throughput between memories. Data set of D8, D16 and D32 were read and written from/to VMEbus at A16, A24 and A32 address space to examine the address space dependency and the data size dependency. There is no address space dependency. Figure 1 shows the measured data of the data transfer speed. The HP743rt is about 30% faster than the IA32 system in any case. The typical value of the measurement is 1.2 Mbytes/sec at D8 on the HP743rt.

Table 2: Interrupt response time with CPU and bus architecture dependency

| Platform | Interrupt response time (msec) |
|---|---|
| HP743rt/HP-RT | 16.7 |
| V155/Solaris 7 | 15.9 |
| PCISYS-56A/Solaris 7 | 15.3 |

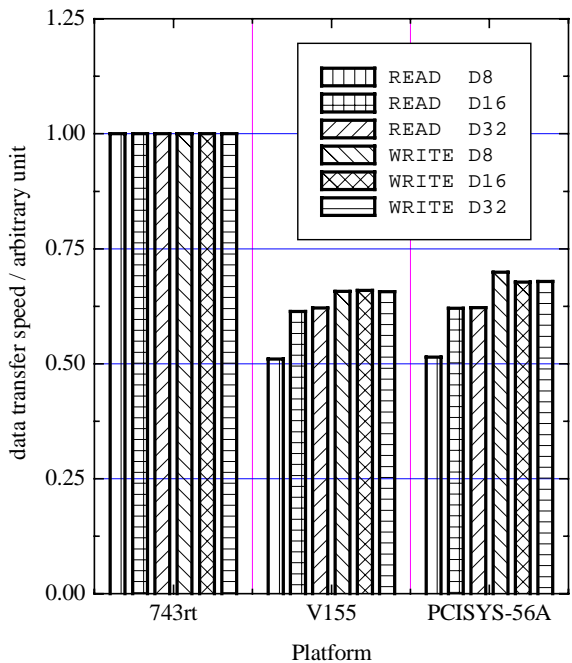

Figure 1: Platform dependency of the data transfer speed with relation to the memory space and the access data size. It normalized to HP9000/743rt to 1.

## 3.3 Context Switching Time

An OS must provide a bounded response time for real-time applications. Context switching time is the overhead when the kernel schedules a process to execution. We measured the context switching time to estimate the guaranteed maximum response time of the system.

Table 3 shows the context switching time of five platforms. Two kinds of measurements were carried out to examine system dependence on CPU load. The first two user processes were running to measure the switching time without any unnecessary load in the single user mode. Next another application with a high CPU load was running in the background. RT-class is used for the measurement in Solaris 7. The results depend on not only the CPU performance but also the scheduling overhead of OS. In comparison of the results between Xycom658 and Advme8001, a difference depends on the processing performance of the CPU. The measured value of the HP743rt is consistent with the data of the technical datasheet from Hewlett-Packard [8]. Linux shows the best performance without any load, but it goes down under the CPU load because the standard Linux cannot assign a fixed priority to the process.

The preemptive kernel, Solaris 7, was slow. This delay might come from preemption. Because Solaris 7 can assign the fixed priority to the process dynamically, the context switching time does not depend on the CPU load.

Table 3: Context switching time with platform dependency and correlation with CPU load.

| Context switching time ($\mu$sec) | | |
|---|---|---|
| HP743rt/HP-RT | No load | 194.3 (SD=49.9) |
| | CPU load | 193.0 (SD=23.5) |
| Xycom658/Solairs7 (RT-class) | No load | 253.5 (SD=165.3) |
| | CPU load | 250.6 (SD=175.7) |
| Advme8001/Solaris7 (RT-class) | No load | 90.7 (SD=55.3) |
| | CPU load | 90.3 (SD=58.6) |
| Xycom658/Linux (2.2.9) | No load | 58.5 (SD=76.7) |
| | CPU load | 10050 (SD=16030) |
| Advme8001/Linux (2.2.9) | No load | 21.9 (SD=15.8) |
| | CPU load | 7020 (SD=1204) |

## 4 CONCLUSIONS

We measured the interrupt response time, data transfer speed and the context switching time as real-time performance. The data transfer speed on the IA32 system is a little slower than the HP743rt system, however it is tolerable in our system. The decreasing of the data transfer speed comes from latency of the PCI-VMEbus bridge. The Linux system shows the fastest context switching time. However the scheduler of the standard Linux kernel changes process priorities dynamically. It worsens context switching time at high CPU load. We did not choose the standard Linux because of its lack of fixed priority control that is essentially needed for our purpose.

Because of the improvement of the CPU performance on the IA32 system, we are able to obtain a good enough performance on Solaris wholly. Finally we decided on two kinds of CPU boards, Xycom XVME658 and Advanet Advme8001 with a fan-less configuration.

Recently many vendors provide real-time extensions of Linux that can control process priorities. We will test Linux with real-time extensions in the future.